\documentclass[amsmath,showkeys, amssymb,a4paper,11pt]{revtex4}
\usepackage{latexsym}
\usepackage[dvipdfm,colorlinks=true,linkcolor=cyan,anchorcolor=cyan,citecolor=cyan]{hyperref}
\usepackage[hmargin={2.54cm,2.54cm},vmargin={3.17cm,3.17cm}]{geometry}
\usepackage{float}
\usepackage{graphicx,amsfonts,amsmath,amssymb,amsthm,amscd}
\usepackage{epsfig}
\usepackage{url}
\usepackage{subfigure}
\usepackage{bm}
\usepackage{xr}
\usepackage{indentfirst}
\usepackage{autobreak}

\def \w{\omega}
\def \b{\widetilde{b}}
\def \a{\widetilde{a}}
\def \C{\widetilde{C}}
\def \B{\widetilde{B}}
\def \T{\widetilde{T}}
\def \S{\widetilde{S}}

\def \Y{\widetilde{Y}}
\def\d{\mathrm{d}}

\newcommand{\Eref}[1]{Eq.\,(\ref{#1})}
\newcommand{\Fref}[1]{Fig.\,\ref{#1}}

\graphicspath{{fig/}}

\date{\today}

\begin{document}

\title{
  \bf Mean velocity and effective diffusion constant for translocation of biopolymer chains across membrane
}

\author{Xining Xu and Yunxin Zhang} \email[Email: ]{xyz@fudan.edu.cn}
\affiliation{Laboratory of Mathematics for Nonlinear Science, Shanghai Key Laboratory for Contemporary Applied Mathematics, Centre for Computational Systems Biology, School of Mathematical Sciences, Fudan University, Shanghai 200433, China.}

\begin{abstract}
Chaperone-assisted translocation through a nanopore embedded in membrane holds a prominent role in the transport of biopolymers.  Inspired by classical {\it Brownian ratchet}, we develop a theoretical framework characterizing such translocation process through a master equation approach. In this framework, the polymer chain, provided with reversible binding of chaperones, undergoes forward/backward diffusion, which is rectified by chaperones. We drop the assumption of timescale separation and keep the length of a polymer chain finite, both of which happen to be the key points in most of the previous studies. Our framework makes it accessible to derive analytical expressions for mean translocation velocity and effective diffusion constant in stationary state, which is the basis of a comprehensive understanding towards the dynamics of such process. Generally, the translocation of polymer chain across membrane consists of three subprocesses: initiation, termination, and translocation of the main body part of a polymer chain, where the translocation of the main body part depends on the binding/unbinding kinetics of chaperones. That is the main concern of this study. Our results show that the increase of forward/backward diffusion rate of a polymer chain and the binding/unbinding ratio of chaperones both raise the mean translocation velocity of a polymer chain, and roughly speaking, the dependence of effective diffusion constant on these two factors achieves similar behavior.
\end{abstract}

\keywords{rectified diffusion, mean velocity, effective diffusion constant, Brownian ratchet}

\maketitle

\section{Introduction}
Translocation across membranes is ubiquitous during or after the synthesis of biopolymers.  A typical example is the precursor proteins destined for the matrix of mitochondria or the endoplasmic reticulum \cite{Walfter2002The, Rapoport2007Protein, Neupert2015A}.  Additional examples, including the export of RNA across nuclear membranes in eukaryotic cells \cite{Santos1998Nuclear, Schatz1996Common} and the viral injection of DNA into a host \cite{MolecularBiologyCell, Liu2014viral}, both involve the translocation through  nanoscopic pores embedded in biological membranes. Besides, new evidence indicates that the uptake of long DNA molecule concerns linear passage through nuclear pore complex \cite{Salman2001Kinetics}. Despite the lack of a membrane-bound nucleus or any membrane-bound organelle, the DNA of bacteria travels across cell envelope through narrow constrictions during the process of horizontal gene transfer \cite{Allemand2010Bacterial, Burton2010Membrane,Chen2005Ins}, such as transformation \cite{Chen2004DNA}, which is responsible for the adaptive evolution of bacteria.  Similar transport mechanisms have also been reported in the biotechnology of drug delivery \cite{Holowka2007Polyarginine} and rapid
DNA sequencing as well \cite{Turner2002Confinement}.

In recent decades, wide attention has been attracted to understanding the mechanism of such crucial biological process.  Pioneering work has established various mechanistic understandings of what induces the directional movement of biopolymers through narrow pores.  One important driving force is external electric field across the membrane as demonstrated by experiments {\it in vitro}  \cite{Kasianowicz1996Chara, Meller2000Rapid} and extensive theoretical analysis \cite{Sung1996Polymer, Meller2001Voltage, Corsi2006Field}.
Considering there is usually no strong enough electric field in living cells, the other mechanism, {\it Brownian ratchet}, suggests pure random thermal diffusion is rectified by chemical asymmetry between the {\it cis} and {\it trans} sides of the membrane, giving rise to a directional motion \cite{Simon1992What, Peskin1993Cellular}. That is exactly the main interest of this paper. A typical case, particularly for proteins, is that the binding of chaperones to the translocating polymer on the {\it trans} side of a membrane prohibits the polymer chain's backward diffusion to the {\it cis} side, thus leading to the directional translocation \cite{Elston2000Models, Matlack1998Protein, Walfter2002The}. This mechanism appears to be confirmed on the basis of empirical data for the transport of prepro-$\alpha$ factor into the lumen of endoplasmatic reticulum, where prepro-$\alpha$-factor is regarded as protein and BiP serves as a chaperone molecule \cite{Matlack1999BiP}.  Recent progress on the kinetic of uptake of DNA also provides evidence for analogous mechanism \cite{Hepp2016Kinetics}.

Quantitative investigations devoted to chaperone-assisted transcription mainly focus on the dynamic properties of the system, with theoretical explorations set up mostly through a continuous space and time description \cite{Simon1992What,Peskin1993Cellular,Roya2003What}. Such continuum models can work effectively in the limit that chaperones are of much lager size compared with the translocating polymer. However, this may become unrealistic in living cells.

Other studies contribute to translocation process of such structure usually through discrete master equations. Early results based on one-dimensional master equations are derived on the simple assumption that both the attachment and the detachment of corresponding chaperones are much faster than the translocation rate, leaving the state of individual sites drawn from the stationary distribution \cite{Ambj2004Chaperone, Ambj2005Directed}. Dropping this approximation, a more general discrete model is then proposed in \cite{D2007Exact}, where the translocating polymer is represented by a one-dimensional lattice of infinite length and an analytical expression for the transcription velocity is obtained mathematically. In \cite{Krapivsky2010Fluctuations}, extensive analysis was carried out on this model to determine diffusion constant and even higher cumulants of the translocated length, while the same process is discussed from the viewpoint of first-passage time  in \cite{Abdolvahab2011First}. All these three models enjoy a common feature, that is they always keep track of the state of each binding site having been translocated to the target region. This approach, although ideal, seems not feasible, especially for long polymer chains, since the space of states shall experience an explosion as the number of binding sites grows. In this sense, such model makes a difference merely on the numerical simulation side, even if explicit expressions for some dynamic quantities can be obtained.  To avoid this issue, a fresh study on this problem set up a modified translocation ratchet model involving the state of only a few binding sites \cite{Uhl2018Force}.  Although a model is proposed dropping the key assumption of timescale separation between the binding/unbinding process of chaperones and the forward/backward diffusion process, analytical results become no longer accessible. Meanwhile, the length of a particular polymer chain is always assumed to be infinite on both side of the membrane in this model, which is not quite appropriate.

In this paper, based on the classical {\it Brownian ratchet}, a general framework describing the chaperone-assisted translocation of a polymer chain across membrane is provided. Since the binding/unbinding kinetics of chaperones might have a sizable effect on the dynamics of translocation process, as addressed in previous work \cite{D2007Exact,Krapivsky2010Fluctuations},  the state of binding sites locating just behind the membrane, {\it i.e.}, whether it is bound with a chaperone, is taken into account explicitly in our model. Moreover, the usual assumption of infinite length of polymer chain is also removed. That is to say, the number of binding sites of a particular polymer chain,  or say the length of polymer chain, is assumed to be finite in our model. Through a  scheme similar to that in \cite{Derrida1983Velocity}, the mean velocity and effective diffusion constant of polymer translocation are obtained explicitly.

This paper is organized as follows. On the concept of {\it Brownian ratchet}, a model describing the process of polymer translocation across membrane is introduced in \autoref{sec2}. How this model works will be explained in detail with coupled master equations. In \autoref{sec3}, we deduce analytical expressions for mean translocation velocity and effective diffusion constant. To show basic properties of the translocation process, numerical results are presented in \autoref{sec4}. Finally, we summary our results  briefly in \autoref{sec5}.

\section{Theoretical description of polymer translocation across membrane through a nanopore}\label{sec2}
During the translocation across membrane, a polymer chain moves stochastically to either side of the pore embedded in the membrane, which is the only possible gate that polymers can finally climb out. Chaperones are assumed to exist in the target region only. Once the binding site of polymer chain adjacent to the pore is occupied with a chaperone, the polymer chain will then be prevented to move backward through the pore due to the large relative size of a chaperone with respect to the diameter of the narrow pore.

A polymer chain of finite length is represented as a one-dimensional lattice with $N$ binding sites, labeled as $1, 2,\cdots, N$. For convenience, the distance between each two neighboring binding sites is normalized to one.  If we take the polymer chain as the reference frame, the process of translocation of a polymer chain across membrane through a pore can be regarded as the membrane's stochastic motion along a one-dimensional lattice. Besides, we represent the membrane as a wall perpendicular to the polymer chain for simplicity. If the wall is standing at site $k$ and the previous site, labeled as $k-1$, remains unoccupied, the wall may hop stochastically either forward to the next site $k+1$ with rate $\w_{f_0}$, or backward to site $k-1$ with rate $\w_{b}$. Otherwise, if the site $k-1$ is occupied with a chaperone, the wall will have no choice but to hop forward stochastically by one step and the corresponding rate is denoted as $\w_{f_1}$.  In summary, the stochastic motion of membrane along polymer chain can be divided into two categories, one is the usual diffusion process with possible forward and backward hopping and the other is the rectified diffusion with possible forward hopping alone, since the backward hopping is blocked by chaperones. In this sense, our model can be regarded as a {\it Brownian ratchet} translocation model.

We assume that chaperones can stochastically bind to any unoccupied site behind the wall with rate $\w_a$. At the same time, any attached chaperone can detach from its binding sites at random as well, where the detachment rate is denoted as $\w_d$. It's easy to find that the dynamics of a translocation process depend on the states of all the binding sites of the polymer chain behind the wall. An ideal model, therefore, should include the states of all these binding sites, {\it i.e.}, whether it is occupied by some chaperone or not. However, keeping all these states in the model explicitly will make further theoretical analysis inaccessible, especially for long polymer chains, since the state space grows exponentially with the number of binding sites.

Taking this into consideration, we assume that the binding sites two units or more away from the wall have sufficient time to reach equilibrium, and the probability of staying unoccupied thereby comes to $q=\w_d /(\w_a +\w_d)$. Actually, the binding/unbinding kinetics of chaperones is much faster than the forward/backward diffusion process of a polymer chain's translocation across the membrane. The ratio of their timescales lies around $1/300$ \cite{Ulrich2002Physical}, which makes the assumption plausible therefore.

\begin{figure}[h]
	\centering
	\includegraphics[width=0.9\textwidth]{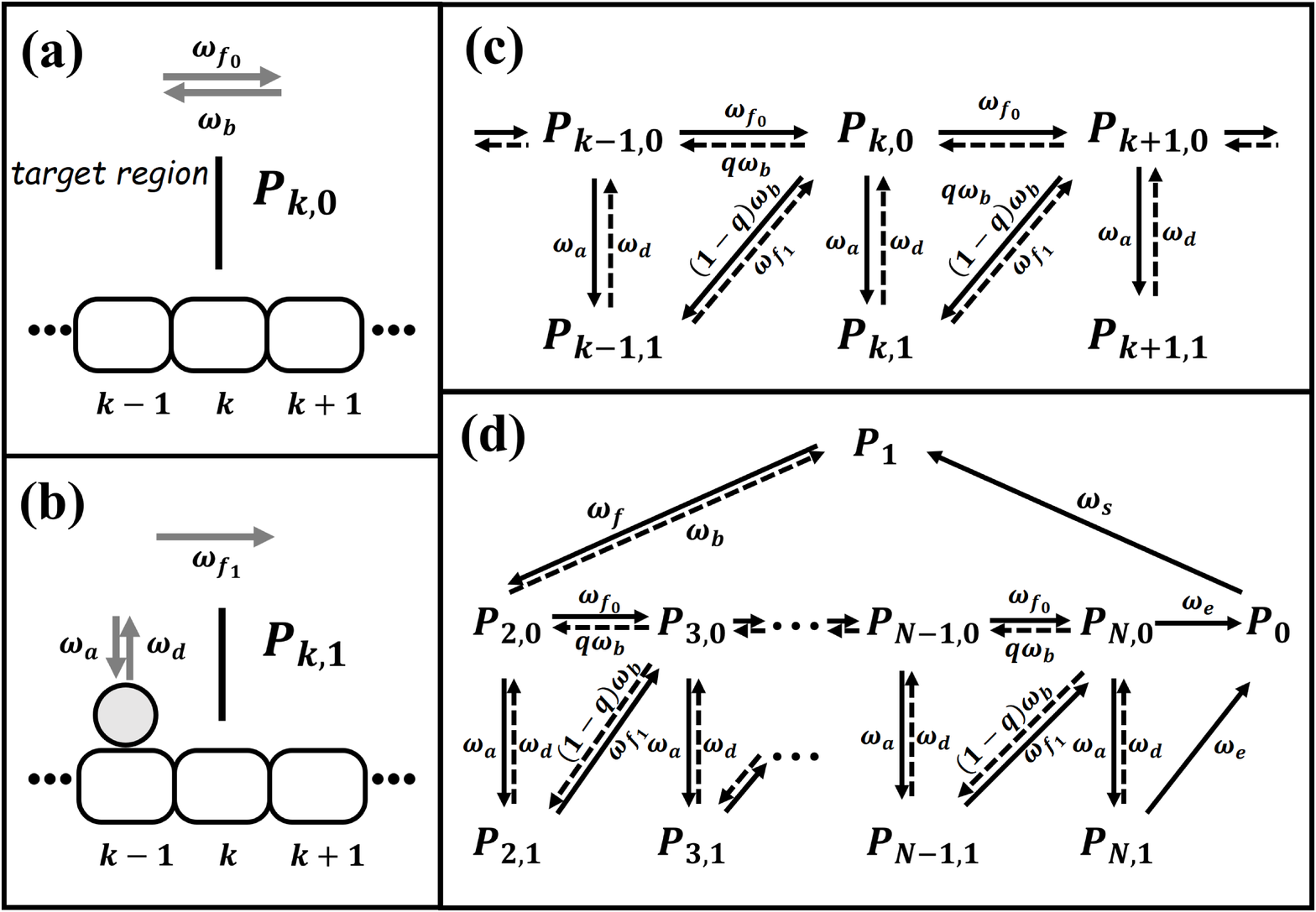}
	\caption{Schematic depiction of the chaperone-assisted translocation of polymer chain across membrane. (a) and (b) depict the two possible states included in our model, that is whether the last site transferred is unoccupied or occupied with a chaperone. The corresponding probabilities are denoted as $P_{k,0}$ and $P_{k,1}$ respectively. Here, the linear polymer molecule is represented by a one-dimension lattice, the bar stands for the moving wall (membrane) and the circle is for a chaperone. Target region of polymer chain is on the left side of the wall. (c) provides a detailed description of possible transitions around a main body site $k$ ($2\le k\le N-1$) and related transition rates included in master equations, while (d) shows the transitions around boundary sites $1$ and $N$. }
	\label{Schematic}
\end{figure}

Our model characterizing the stochastic translocation of a polymer chain across membrane, or equivalently the stochastic motion of a wall along a one-dimensional lattice with length $N$ is depicted in \Fref{Schematic}. For $k = 2,\dots, N$, we denote $P_{k,0}$ and $P_{k,1}$ as the probability that the wall stands at site $k$ with the previous site $k-1$ unoccupied or occupied with a chaperone respectively, see \Fref{Schematic}(a, b). The first site of a polymer chain is labeled as $1$, and the corresponding probability that the wall stands at this site, site $1$, is denoted as $P_1$. Additionally, we denote $P_0$ as the probability that none of the sites of a polymer chain is passing through the pore, {\it i.e.}, the nanopore of the membrane is vacant. In other words, $P_0$ is the probability that a polymer chain has just accomplished its translocation, and a new translocation process has not started yet.

Since chaperones just exist in the target region,  only the sites behind the wall are permitted to be bound with chaperones. Then, the wall, standing at site $k$ currently, can always move forward stochastically to the next site $k+1$, while it gets quite different for backward motion. If the site $k-1$ is unoccupied, the wall at site $k$ can also stochastically move backward to site $k-1$ with rate $\w_{b}$. However, if the site $k-1$ is occupied with a chaperone, the backward motion of the wall at site $k$ will be blocked. That is to say, the forward motion of the wall standing at site $k$ may depend on the state of site $k-1$. To state a general case, the forward hopping rate of the wall is denoted differently, where $\w_{f_0}$ and $\w_{f_1}$ match the cases that site $k-1$ stays in unoccupied and occupied state.


To picture the site-by-site translocation process, we can still imagine a wall standing at some main body site, labeled as $k$. As described in the beginning of this section,the probability that site $k-2$ stays in unoccupied state is $q$, with this site reaching the equilibrium of binding/unbinding kinetics. Based on sustained translocation process, we establish a one-dimensional system of period $l$, where one period indicates a complete transfer of one identical polymer chain.  Then, master equations of probabilities $P_{k,0}$ and $P_{k,1}$ in the main body of a lattice, {\it i.e.}, $k = 3,\cdots, N-1$ are given by
\begin{align}
\label{MEk0}
\frac{ \partial P_{k,0} (l, t)} {\partial t }  =&
q \w_{b} P_{k+1,0}(l, t)+ \w_{f_0}P_{k-1,0}(l, t) + \w_{f_1}P_{k-1,1}(l, t) \nonumber \\
&+ \w_d P_{k, 1}(l, t)-(\w_b + \w_a +\w_{f_0}) P_{k,0}(l, t)  \, , \\
\label{MEk1}
\frac{ \partial P_{k,1} (l, t)} {\partial t }  =& (1-q) \w_b P_{k+1,0}(l, t) + \w_a P_{k,0}(l, t) -(\w_d + \w_{f_1}) P_{k,1}(l, t) \, .
\end{align}
See \Fref{Schematic}(c) for the schematic depiction.

Since site $1$ is the starting point of translocation process (see \Fref{Schematic}(d)), there is no lattice site behind the wall if it stands at site $1$, indicating that no chaperone can be attached to the polymer chain when $k=1$. Thus, the forward hopping rate of the wall at site $k=1$ may be different from both $\w_{f_0}$ and $\w_{f_1}$. Therefore, a different notation $ \w_{f}$ is used to denote the forward hopping rate of the wall for $k=1$.  As schematically depicted in \Fref{Schematic}(d), \Eref{MEk1} still holds for $k = 2$ while \Eref{MEk0} should be modified as
\begin{equation}
\label{ME20}
\frac{ \partial P_{2,0} (l, t)} {\partial t }  =
q \w_b P_{3,0}(l, t)+ \w_{f}P_{1}(l, t) + \w_d P_{2, 1}(l, t)-(\w_b+ \w_a +\w_{f_0}) P_{2,0}(l, t)\, .
\end{equation}

When it comes to the boundary sites, things get different. As discussed above, we use $P_0$ to denote the probability that the system stays in the interval between two translocation processes of separate polymer chains, and in this state, the wall can be regarded to stand at a fictitious site $0$. Translocation starts with rate  $\w_{s}$ when the wall binds to the initial lattice site $1$ of a polymer chain, or we say the wall jumps from the fictitious site $0$ to site $1$ with rate $\w_S$. Similarly, it is when the wall hops forward from the terminal site $N$ to the fictitious site $0$ with rate $\w_{e}$ that the translocation of a polymer chain comes to an end. Also see \Fref{Schematic}(d) for a schematic depiction. Based on this scenario,  the master equation for probability $P_{1} $ is as follows,
\begin{equation}
\label{ME1}
\frac{ \partial P_{1} (l, t)} {\partial t }  = \w_s P_0(l,t) +\w_b P_{2,0}(l,t) -\w_f P_{1}(l,t) \,.
\end{equation}
Similarly, probabilities $P_{N,0}$ and $P_{N,1}$  are governed by
\begin{align}\label{MEn}
\frac{\partial P_{N,0}(l, t)}{\partial t} &= \w_{f_0} P_{N-1,0}(l, t) + \w_{f_1} P_{N-1,1}(l, t) +\w_{d}P_{N,1}
- (\w_b + \w_a +\w_e) P_{N,0}(l, t) \, ,\\
 \label{MEn1}
\frac{\partial P_{N,1}(l, t)}{\partial t} &= \w_a P_{N,0}- (\w_d +\w_e) P_{N,1}(l, t) \, .
\end{align}

Once a polymer chain has been transferred to the target region completely, the translocation process enters into a pause state, where the wall will wait until it walks onto another polymer chain, that is when another polymer chain enters into the nanopore. With \Fref{Schematic}(d), the probability $P_0$  satisfies
\begin{equation} \label{ME0}
\frac{ \partial P_{0} (l, t)} {\partial t } = \w_e [ P_{N,0}(l-1, t) + P_{N,1}(l-1, t) ] - \w_s P_0(l,t) \, .
\end{equation}

Hence, we have set up a system of $2N$ states, where the translocation of a polymer chain across a nanopore of membrane is mapped into the motion of a wall through a one-dimensional lattice in terms of relative motion.

\section{Expressions for Mean Velocity and Effective Diffusion Constant}\label{sec3}
In this section, we will follow the main idea of \cite{Derrida1983Velocity,Kolomeisky2005Dynamic} to realize analytical expressions for mean velocity and effective diffusion constant of translocation of polymer chains across membrane in the stationary-state limit {\it i.e.}, $t \rightarrow \infty$.  We begin with a general expression for the mean location of the wall, which is
\begin{equation}\label{Esite}
\langle x(t)  \rangle = \sum_{l = - \infty} ^{+\infty} \sum_{k=0}^{N} (k+Nl) P_k(l,t)  \, .
\end{equation}
That is to say, if we assume the length of each biopolymer chain equals $N$, each translocation period will add $N$ units to the accumulated distance that the wall travels, since hopping away from a polymer chain (translocation {\it termination}, from site $N$ to fictitious site $0$), or stepping into a new period (translocation {\it initiation}, from fictitious site $0$ to site $1$), does not increase the total distance that the wall travels.  Note that, in \Eref{Esite} $P_k(l,t) :=P_{k,0}(l,t) +P_{k,1}(l,t)$ for $2\le k\le N$, is the probability that the wall stands at site $k$ of the $l-$th polymer at time $t$(see \Fref{Schematic}). We now define a few auxiliary functions for each state concerned, which are shown as
 \begin{equation} \label{def1}
B_k(t) = \sum_{l = -\infty}^{+\infty} P_{k,0}(l, t),
\quad
C_k(t) = \sum_{l = -\infty}^{+\infty} (k+Nl) P_{k,0}(l, t) \, ,
\end{equation}
\begin{equation} \label{def2}
\B_k(t) = \sum_{l = -\infty}^{+\infty} P_{k,1}(l, t) \,,
\quad
\C_k(t) = \sum_{l = -\infty}^{+\infty} (k+Nl) P_{k,1}(l, t) \,.
\end{equation}
Notably, functions in \Eref{def1} are defined for index $k$ such that $0\leqslant k \leqslant N$ whereas those in \Eref{def2} are defined for  $2\leqslant k \leqslant N$ only. Here, for convenience, we assume $P_{k,0}(l, t)=P_k(l, t)$ for $k=0, 1$ in \Eref{def1}.
Generalizing the original Derrida's method \cite{Derrida1983Velocity,Kolomeisky2005Dynamic}, we propose the following ansatz in the stationary-state limit when time goes infinity.
\begin{equation} \label{blim}
B_k(t) \rightarrow b_k, \quad \B_k(t)\rightarrow \b_k \, ,
\end{equation}
\begin{equation} \label{clim}
 C_k(t) \rightarrow a_k t + T_k , \quad  \C_k(t) \rightarrow \a_k t +\T_k \, ,
\end{equation}
where the range of index $k$ is the same as that in Eqs.\,(\ref{def1}, \ref{def2}).  Obviously, all these $b_k\,$s together with $\b_k\,$s are governed by the normalizing condition
\begin{equation}\label{normal}
b_0 +b_1 +\sum_{k=2}^{N}(b_k + \b_k) =1 \, ,
\end{equation}
since $b_k$ (or $\b_k$) gives the probability of finding the wall at a specific state at large time $t$.

In the stationary-state limit, which means $B_k(t)  \rightarrow b_k$,
$\B_k \rightarrow \b_k$ and $\d B_k/\d t = 0$, $\d \B_k/\d t = 0$ when  $t\rightarrow \infty$, master equations given in \autoref{sec2} are transformed into
\begin{equation}\label{eqbk}
\begin{split}
0=& - \w_s b_0 + \w_e(b_N +\b_N)  \, ,\\
0=& \w_s b_0 - \w_f b_1 + \w_b b_2  \, ,\\
0=& \w_f b_1-(\w_b+\w_{f_0}+\w_a)b_2+ \w_d \b_2 + q\w_b b_3 \, ,\\
0=& \w_{f_0}b_{k-1}+\w_{f_1} \b_{k-1}-( \w_a+\w_{f_0}+\w_b) b_k +\w_d \b_k +q\w_b b_{k+1}\, , \quad (k = 3,\cdots,N-1)\, ,\\
0=& \w_{f_0}b_{N-1}+\w_{f_1} \b_{N-1}-( \w_a +\w_e + \w_b) b_N +\w_d \b_N \,,\\
0=&\w_a b_k - (\w_d +\w_{f_1})\b_k + (1-q)\w_b b_{k+1} \, , \quad (k = 2,3,\cdots,N-1) \, ,\\
0=&\w_a b_N - (\w_d +\w_e)\b_N \,.
\end{split}
\end{equation}
The last two equations yield
\begin{equation}\label{bk1}
\b_k =\left\{
\begin{split}
& (\w_d+\w_e)^{-1}\w_a b_k \, ,  \quad (k =N)  \, ,\\
& (\w_d + \w_{f_1}) ^{-1}[ \w_a b_{k} + (1-q) \w_b b_{k+1} ] \, ,  \quad (k = 2,3,\cdots,N-1) \, .
\end{split}
\right.
\end{equation}
The substitution of \Eref{bk1} into \Eref{eqbk} leads to a recurrence
\begin{equation}\label{recurrence}
Rb_{N} - S b_{N-1}  = \cdots =  R b_{k} - S b_{k-1} = \cdots = Rb_3 - Sb_2  = \w_b b_2 - \w_f b_1 = -\w_s b_0  \, ,
\end{equation}
where
\begin{equation*}
R = \w_b \frac{\w_d+q\w_{f_1}}{\w_d+\w_{f_1}}  \quad \text{and}  \quad S = \w_{f_0} + \frac{\w_a \w_{f_1}}{\w_d+\w_{f_1}} \,.
\end{equation*}
\Eref{recurrence} means that the probability fluxes between each two neighboring {\it effective} sites $k-1$ and $k$ remain constant in steady state.

Considering the stationary state for boundary sites, {\it i.e.} the first equation and the last equation in \Eref{eqbk}, we have
\begin{equation}\label{bn}
b_N =\left[ \w_e +\frac{\w_a \w_e}{\w_d+\w_e}\right]^{-1} \w_s b_0  \equiv K \w_s b_0\,.
\end{equation}

The recurrence given in \Eref{recurrence} along with \Eref{bn} yields a general expression for $b_k$,
\begin{equation}\label{bkg}
\begin{split}
b_k &= \left[
\S^{k-N}[K-(S-R)^{-1}]+ (S-R)^{-1}
\right]\w_s b_0 \, , \quad (k =2, \cdots, N) \, , \\
b_1 &= \w_f  ^{-1}( \w_b b_2 + \w_s b_0 )\, ,
\end{split}
\end{equation}
where $\S \equiv S/R$.
Thus, all of $b_k\,$s and  $\b_k\,$s are determined by $\w_s b_0$, which is the probability flux exactly. See \Eref{recurrence}.
With the normalizing condition given in \Eref{normal}, we obtain
\begin{equation}\label{b0}
\w_s b_0 = \left[
W_1 + W_2 \S^{2-N} + W_3 (N-1)
\right]^{-1} \,,
\end{equation}
where
\begin{equation*}
W_1 = \frac{1}{\w_s} + \frac{1}{\w_f}
+K(\frac{\w_a}{\w_d+\w_e} -\frac{\w_a}{\w_d+\w_{f_1}})
+\frac{U}{S-R}
+M\frac{S}{S-R}(K-\frac{1}{S-R}),
\end{equation*}
\begin{equation*}
W_2 =\left(K-\frac{1}{S-R}\right)
\left( U-\frac{M}{\S-1}\right)
\quad \text{and} \quad
W_3 = \frac{M}{S-R}\, ,
\end{equation*}
with
\[ M = 1+ \frac{\w_a}{\w_d+\w_{f_1}} +\frac{(1-q)\w_{b}}{\w_d+\w_{f_1}}
\quad  \text{and} \quad
U=\frac{\w_b}{\w_f} - \frac{(1-q)\w_b}{\w_d+\w_{f_1}}\,.
 \]
So far, $b_0$ has been obtained explicitly as a function of all constant rates.  So do $b_k\,$s and $\b_k\,$s. See Eqs. (\ref{bkg}, \ref{b0}) and \Eref{bk1}.

Let's return to the ansatz given in \Eref{clim} to derive explicit expressions for $a_k\,$s, $\a_k\,$s, $T_k\,$s and $\T_k\,$s.
With the definitions of $C_k(t)$ and $\C_k(t)$ (see \Eref{def1} and \Eref{def2}) as well as the master equations presented in \autoref{sec2},  it is easy to show that $C_k(t)$ and $\C_k(t)$ satisfy
\begin{equation}\label{C}
	\begin{aligned}
		\frac{ \d C_0} {\d t }   (t)  = &
				-  \w_{s} C_{0} + \w_e (C_{N} + \C_{N})   \, , \\
		\frac{ \d C_1} {\d t }   (t)  = &
 			\w_{s} C_{0} - \w_{f} C_{1} +\w_b C_{2} +\w_{s} B_{0} -\w_b B_{2} \, , \\
		\frac{ \d C_2} {\d t }   (t)  = &
			\w_{f} C_{1} -(\w_b + \w_a +\w_{f_0}) C_{2}  + \w_d \C_{2}  +q \w_b C_{3}
			+\w_{f} B_{1} - q\w_b B_{3} \, , \\
		\frac{ \d C_k} {\d t }   (t)  =  &
		 	\w_{f_0} C_{k-1} + \w_{f_1} \C_{k-1}  -(\w_b + \w_a +\w_{f_0}) C_{k} + \w_d \C_{k}+q \w_b C_{k+1} \\
		 	& +\w_{f_0} B_{k-1} + \w_{f_1}\B_{k-1} -q\w_b B_{k+1}\,, \quad (k= 3,\dots, N-1) \, ,\\
		\frac{ \d C_N} {\d t }   (t)  =  &
			\w_{f_0} C_{N-1} + \w_{f_1} \C_{N-1} -(\w_b + \w_a +\w_{e}) C_{N} + \w_d \C_{N} +\w_{f_0} B_{N-1} + \w_{f_1} \B_{N-1}\, , \\
		\frac{ \d \C_{k}} {d t}  (t) =& \w_a C_{k} -(\w_d +\w_{f_1}+) \C_{k} + (1-q) \w_b C_{k+1}
		- (1-q) \w_b B_{k+1}\,, \quad (k= 2,\dots, N-1) \, ,\\
		\frac{ \d \C_{N}} {d t}  (t) =&   \w_a C_{N} -(\w_d +\w_{e}) \C_{N}  \,.\\
 	\end{aligned}
\end{equation}

Let time $t\rightarrow \infty$ and focus on the terms proportional to time $t$. Then, with of the help of \Eref{clim} and \Eref{C}, we get the following expressions,
\begin{equation}\label{a}
	Ra_{N} - Sa_{N-1} = \cdots = Ra_{k} - Sa_{k-1} =\cdots =\w_b a_2 -\w_f a_1 = -\w_s a_0 \, ,
\end{equation}
\begin{equation}\label{ak}
	\a_k =\left\{
	\begin{split}
	 & (\w_d+\w_e)^{-1}\w_a a_k \,, \quad (k=N) \, , \\
	 & (\w_d + \w_{f_1}) ^{-1}[ \w_a a_{k} + (1-q) \w_b a_{k+1} ] \, , \quad  (k=2,3,\cdots, N-1) \, .
	\end{split}
	\right.
 \end{equation}

Compare Eqs. (\ref{a}, \ref{ak}) with Eqs. (\ref{recurrence}, \ref{bk1}), and one can immediately conclude that
\begin{equation}\label{abk}
	a_k = Ab_k \,, \quad \a_k = A\b_k\, ,
\end{equation}
and
\begin{equation}\label{eq29}
	a_0 + a_1 +\sum_{k=2}^N(a_k+\a_k) =A[b_0 + b_1 +\sum_{k=2}^N(b_k+\b_k)] = A  \, ,
\end{equation}
where $A$ is a constant.
Sum up all of the equations in the stationary-state limit of \Eref{C} and we have
\begin{equation} \label{coeff}
	\begin{split}
		 A &= \w_s b_0 +\w_f b_1 + \w_{f_0} \sum_{k=2}^{N-1} b_k+\w_{f_1}\sum_{k=2}^{N-1} \b_k-\w_b\sum_{k=2}^{N}b_k\\
		 &=\w_s b_0 + \w_b b_1 - \w_f b_2 + \sum_{k=2}^{N-1}(Sb_k -Rb_{k+1}) \\
		 &= N\w_s b_0\,,
	\end{split}
\end{equation}
where the last equation follows the recurrence in \Eref{recurrence} and $b_0$ is given in \Eref{b0}. Therefore, constant $A$ is determined by all of the constant transition rates. Explicit expressions for $a_k\,$s and $\a_k\,$s can also be obtained via Eqs.\,(\ref{bk1},  \ref{bkg}, \ref{b0}, \ref{abk}, \ref{coeff}).

Next, we begin to derive the expressions for $T_k\,$s and $\T_k\,$s. In the stationary-state limit (see Eqs.\,(\ref{blim}, \ref{clim})), \Eref{C} employs the following expressions,
\begin{equation} \label{T}
	\begin{split}
		a_0= & -  \w_{s} T_{0} + \w_e (T_{N} + \T_{N})   \, , \\
		a_1= &  \w_{s} T_{0} - \w_{f} T_{1} + \w_b T_{2} + \w_{s} b_{0} - \w_b b_{2} \, , \\
		a_2= &  \w_{f} T_{1}   -(\w_b + \w_a +\w_{f_0})T_{2}  + \w_d \T_{2} + q \w_b T_{3} +\w_{f} b_{1} - q\w_b b_{3} \, , \\
		a_k = & \w_{f_0} T_{k-1} + \w_{f_1} \T_{k-1}-(\w_b + \w_a +\w_{f_0}) T_{k} + \w_d \T_{k}+q \w_b T_{k+1}\\
		 & +\w_{f_0} b_{k-1} + \w_{f_1} \b_{k-1} - q\w_b b_{k+1}  \,, \quad (k= 3,\dots, N-1) \, ,\\
		a_N = &	\w_{f_0} T_{N-1} + \w_{f_1} \T_{N-1} -(\w_b + \w_a +\w_{e}) T_{N}  + \w_d \T_{N} +\w_{f_0} b_{N-1} + \w_{f_1} \b_{N-1}\, , \\
		\a_k = &  \w_a T_{k} -(\w_d +\w_{f_1}) \T_{k}+ (1-q) \w_b T_{k+1}- (1-q) \w_b b_{k+1} \quad (k= 2,\dots, N-1) \, ,\\
		\a_N = &   \w_a T_{N} -(\w_d +\w_{e}) \T_{N}  \,.
	\end{split}
\end{equation}
Here, in the stationary-state limit, attention is paid to the terms independent of time $t$ and thus \Eref{C} reduces to \Eref{T}.

With some new auxiliary quantities
\[ y_1  \equiv \w_f T_1-\w_b T_2 \quad \text{and} \quad y_k  \equiv ST_k - R T_{k+1} \quad
(k = 2, \cdots, N-1) \,,\]
\Eref{T} can be transformed into a recurrence, that is
\begin{equation} \label{recuy}
	y_k - y_{k-1} = Q_{k} \, ,
\end{equation}
where $k = 2,\cdots,N-1$. The first term $y_1$ comes as
\begin{equation} \label{recu1}
	y_1 = \w_s T_0 + Q_1 \, .
\end{equation}
Since we have already obtained the explicit expressions for $b_k\,$s as well as $a_k\,$s, $Q_k$s turn  to known terms, read as
\begin{equation*}
	\begin{split}
		Q_1 &= - a_1 -\w_b b_2 +\w_s b_0  \, ,\\
		Q_2 & = -a_2 -\frac{\w_d}{\w_d +\w_{f_1}}\a_2-Rb_3+\w_f b_1 \, ,\\
		Q_k & = -a_k-\frac{\w_{f_1}}{\w_d +\w_{f_1}} \a_{k-1}-\frac{\w_d}{\w_d +\w_{f_1}}\a_k - R b_{k+1} + S b_{k-1}\, , \quad (k= 3, 4,\cdots,N-1) \, .\\
	\end{split}
\end{equation*}
Through a simple summation, it produces
\begin{equation}\label{ygen}
y_k = \w_s T_0 + r_k \, ,\quad (k = 1, \cdots, N-1) \, ,
\end{equation}
 where
\begin{equation}\label{rk}
	 \begin{split}
		r_1 = & Q_1 =  - a_1 -\w_b b_2 +\w_s b_0 \, , \\
		r_2 =&  \sum_{i = 1}^{2} Q_i =  - \sum_{i=1}^{2} a_i  -\frac{\w_{f_1}}{\w_d +\w_{f_1}} \a_{2} - R b_{3}+ 2\w_s b_0 \, ,\\
		r_k =& \sum_{i = 1}^{k} Q_i =  - \sum_{i=1}^{k} a_i  - \sum_{i=2}^{k-1} \a_i -\frac{\w_{f_1}}{\w_d +\w_{f_1}} \a_{k} - R b_{k+1}+ k\w_s b_0 \,, \quad (k = 3, \cdots, N-1)\,.
	 \end{split}
 \end{equation}
 It means that each $y_k$ consists of two components: one has been determined explicitly and the other depends on the undetermined constant $T_0$. In fact, the same holds for each $T_k$, which can be shown through backward iterations. That is
 \begin{equation}\label{TkT0}
 \begin{split}	
 	T_k &= S^{-1}(RT_{k+1} + y_k) \, \quad  (k = 2, \cdots, N-1)\, , \\
 	T_1 &= \w_f^{-1}(\w_b T_2 +y_2) \, ,
 \end{split}
 \end{equation}
and the first term is
 \begin{equation}\label{TnT0}
T_N  = K(a_0 + \w_s T_0) +(\w_a + \w_d +\w_e)^{-1}\a_N\, ,
 \end{equation}
which is derived from the first and the last equations in \Eref{T}. Obviously, $T_N$, and consequently all $T_k\,$s, are made up of these two components as well, {\it i.e.}, one has been determined explicitly by constant transition rates and the other depends on the undetermined constant $T_0$.

We can write  $T_k = X_k +Y_k$  now, where $X_k$ is proportional to $T_0$ and $Y_k$ is described in some determined factors. Particularly, $ T_0 = X_0$ and $Y_0 = 0$. At the same time, $\T_k$s can be derived from the last two equations in \Eref{T} and they are shown as
 \begin{equation}\label{Tk1}
\T_k = \left\{
\begin{split}
&(\w_d +\w_e)^{-1} [\w_a T_k -\a_k]\, , \quad (k=N) \,,\\
&(\w_b +\w_{f_1})^{-1}[\w_a T_k +(1-q)\w_{f_1}T_{k+1}-\a_k-(1-q) b_{k+1}]\, , \quad  (k= 2,\cdots, N-1)\, .  \\
\end{split}
\right.
\end{equation}
That is to say, each $\T_k$
can also be split into those two parts and then we can write $\T_k = \widetilde{X}_k+ \Y_k$ with $ \widetilde{X}_k$ proportional to $T_0$ and $\Y_k$ determined.

Surprisingly, the undetermined constant $T_0$ will cancel out in the final expressions for the dynamic properties concerned. Keeping this in mind, we first let $T_0=0$ for the calculation of $Y_k$ and show how that works later.

With $T_0 = 0$,  that is when both $X_k$ and $\widetilde{X}_k$ equal zero, we find
\begin{equation}\label{Yk0}
\begin{split}
Y_N &  = Ka_0 +(\w_a + \w_d +\w_e)^{-1}\a_N \,,\\
Y_k &  = \S^{k-N}Y_N + S^{-1}\sum_{i=0}^{N-1-k} \frac{r_{i+k}}{{\S}^{i}} \quad (k = 2, \cdots, N-1)\, ,\\
Y_1 & = \w_f^{-1} [\w_b Y_2 + Q_1]  \, ,\\
Y_0 & = 0 \, ,
\end{split}
\end{equation}
and $\Y_k$ is given by $Y_k$ in the same form as \Eref{Tk1}, {\it i.e.},
\begin{equation}\label{Yk1}
\Y_k = \left\{
\begin{split}
&(\w_d +\w_e)^{-1} [\w_a Y_k -\a_k]\, , \quad (k=N )\, ,\\
&(\w_d +\w_{f_1})^{-1}[\w_a Y_k +(1-q)\w_{f_1}Y_{k+1}-\a_k-(1-q) b_{k+1}]\, , \quad (k= 2,\cdots, N-1) \,.
\end{split}
\right.
\end{equation}

Now, we are fully prepared to derive analytical expressions for dynamic properties of interest, {\it i.e.}, the mean translocation velocity $V$ and effective diffusion constant $D$, which are respectively defined as
\begin{equation}\label{Vdef}
V = \frac{\d \langle x(t)  \rangle}{\d t} \, ,
\end{equation}
and
\begin{equation}\label{Ddef}
2D = \frac{\d }{\d t}\left(
\langle x^2(t)  \rangle -\langle x(t)  \rangle^2
\right)
=  \frac{\d \langle x^2(t)  \rangle }{\d t}
 -2 V \langle x(t) \rangle \,.
\end{equation}

With \Eref{Esite} and the auxiliary functions given in Eqs.\,(\ref{def1}, \ref{def2}),
\begin{equation}
V = \sum_{k=0}^{N}\frac{\d C_k(t)}{\d t} +  \sum_{k=2}^{N}\frac{\d \C_k(t)}{\d t}\,,
\end{equation}
which is exactly the summation of \Eref{C}. Recalling the derivation of constant $A$ (see Eqs.\,(\ref{abk}, \ref{eq29}, \ref{coeff})), as well as the stationary-state assumption as is given in \Eref{clim}, we can easily get the mean velocity in the stationary-state limit as follows,
\begin{equation}\label{V}
V =a_0 + a_1 +\sum_{k=2}^N(a_k+\a_k) = A = N\w_s b_0 \, ,
\end{equation}
where the probability flux $\w_s b_0$ is given by \Eref{b0}.

An explicit expression for effective diffusion constant $D$ requires more detailed analysis. Utilizing the master equations in \autoref{sec2}, the differentiation of $\langle x^2(t)  \rangle $ turns to
\begin{equation}
 \frac{\d \langle x^2(t)  \rangle }{\d t} = \sum_{k=0}^{N}(k+Nl)^2 \frac{\d P_{k}(t)}{\d t} = D_1 + D_2 \, ,
\end{equation}
 where
 \begin{align}\label{D1}
 D_1 = & 2\left[
 \w_s C_0 +\w_f C_1 +\w_{f_0} \sum_{k=2}^{N-1} C_k + \w_{f_1} \sum_{k=2}^{N-1} \C_k  -\w_b\sum_{k=2}^{N} C_k
 \right] \, ,\\
 \label{D2}
 D_2= &\left[
 \w_s B_0 +\w_f B_1 +\w_{f_0} \sum_{k=2}^{N-1} B_k + \w_{f_1} \sum_{k=2}^{N-1} \B_k  +\w_b\sum_{k=2}^{N} B_k
 \right]\, .
 \end{align}
 Note that $P_k(l,t)=P_{k,0}(l,t)+P_{k,1}(l,t)$ for any $l, t$ and $2\le k\le N$ while $P_k(l,t)=P_{k,0}(l,t)$ for $k=0,1$, see \Fref{Schematic}(d).

At the same time, the second part of $2D$ as given in \Eref{Ddef} can be described as
\begin{equation}\label{D3}
D_3 \equiv 2V\langle x(t)  \rangle  = 2A\left[
C_0 + C_1 +\sum_{k = 2}^{N}(C_k + \C_k)
\right]\ \,.
\end{equation}

In the stationary-state limit, we first claim that terms proportional to time $t$ in the expression for $D$ cancel out.
Since $B_k(\B_k)$ tends to be a constant $b_k(\b_k)$ as $t\rightarrow \infty$, we just concentrate on \Eref{D1} and \Eref{D3} to verify it. Replace $C_k(\C_k)$ with $a_k(\a_k)$  in \Eref{D1}, one can easily find that the coefficient of $t$ in $D_1$ comes as $2A\cdot N \w_s b_0 = 2A^2$ in comparison with \Eref{coeff}. Similarly, the coefficient of $t$ in $D_3$ is
$$2A\cdot \left[ a_0+a_1+\sum_{k=2}^N(a_k +\a_k) \right] = 2A^2 \left[b_0+b_1+\sum_{k=2}^N(b_k +\b_k)\right] =2A^2 \,.$$

Then, we focus on the terms proportional to $T_0$ in order to show that the effective diffusion constant  $D$ is independent of them. As mentioned before, each $T_k$ and $\T_k$ can be written as $T_k = X_k + Y_k$ and  $\T_k = \widetilde{X}_k + \Y_k$ respectively. With a glance at \Eref{ygen} and \Eref{TkT0}, we find $X_k$ follows the same recurrence as $b_k$ does. In other words, every $b_k$ in \Eref{recurrence} can be replaced with $X_k$.  So does $\widetilde{X}_k$ with respect to $\b_k$ through  a comparison between \Eref{Tk1} and \Eref{bk1}. Thus, we have
\begin{equation}\label{Xk}
X_k = \frac{T_0}{b_0} b_k
\quad \text{and} \quad
\widetilde{X}_k=\frac{T_0}{b_0} \b_k \,.
\end{equation}

This can be better understood if the procedure of seeking $b_k\,$s and $\b_k\,$s is regarded as solving linear equations $\bm{\Omega}\bm{x} = \bm{0}$ corresponding to \Eref{eqbk}, where $\bm{\Omega}$ is a matrix of  $2N\times2N$ in size.  Since the rank of $\bm{\Omega}$ is $2N-1$, $\bm{b} = (b_0,b_1,b_2,\b_2,\cdots,b_N,\b_N)^\intercal$, what we obtained in the foregoing, is determined on the normalizing condition.  If we denote $\bm{T}$ as the vector $(T_0,T_1,T_2,\T_2,\cdots, T_N,\T_N)^\intercal$, one can find the derivation of $\bm{T}$ is just solving the linear equation $\bm{\Omega}\bm{x} = \bm{Z}$ due to \Eref{T}, where each component in $\bm{Z}$ is determined by $b_k(\b_k)$ and $a_k(\a_k)$. Actually, we have already obtained a solution for this equation, which is $\bm{Y}=(Y_0, Y_1,Y_2,\Y_2,\cdots,Y_N,\Y_N)^\intercal$ with each component given in \Eref{Yk0} and \Eref{Yk1}. Then, a general solution $\bm{T}$ can be written as $\bm{T} = \alpha \bm{b} +  \bm{Y}$. Particularly, the first component in $\bm{Y}$ is zero, {\it i.e.} $Y_0 = 0$. If we regard $T_0$ as a undetermined constant, the undetermined constant should be restricted to $\alpha = T_0/b_0$ in the consideration of the first component in $\bm{T}$, $\bm{Y}$ and $\bm{b}$. \Eref{Xk} holds therefore. That is exactly what we have done.

In this way, considering whether the terms proportional to $T_0$ in \Eref{D1} and \Eref{D3} can cancel out is exactly to examine the performance of $X_k\,$s. It's easy to see such terms in the former equation sum up to $N\w_s T_0$, if we take $X_k$ as a substitution for $C_k$ in \Eref{D1} and compare the coefficients with those in \Eref{coeff}. The latter equation, \Eref{D3}, leads to a summation of $X_k\,$s and $\widetilde{X}_k\,$s, which is $N\w_s T_0$ as well. As a result, terms proportional to $T_0$ cancel out.

Due to these two cancellations, we finally derive the explicit expression for effective diffusion constant
\begin{equation}
\label{D}
\begin{split}
D =&\frac{1}{2} (D_1+ D_2 -D_3)\\
=&-\sum_{k=1}^{N-1}s_k  -\sum_{k=2}^{N-1}\widetilde{s}_k+\frac{V}{2} N -V\left[Y_1+ \sum_{k=2}^{N} (Y_k+\Y_k)\right] \,,
\end{split}
\end{equation}
 where $Y_k$ has already been obtained in \Eref{Yk0}, $\Y_k$ follows \Eref{Yk1}
and
\begin{equation*}
s_k = \sum_{i = 1}^{k} a_i \, , \quad \widetilde{s}_k = \sum_{i=2}^{k}\a_i \,.
\end{equation*}
See Appendix \ref{app} for more details.

\section{Properties of polymer translocation across membrane}\label{sec4}
To further understand the analytical results and get an intuition about the translocation process of polymer chains across membrane,  extensive numerical explorations are carried out according to explicit results given in \Eref{V} and \Eref{D}. Instead of general analysis with respect to each transition rate,  particular attention is paid to a special case where $\w_{f_0} = \w_{f_1} = \w_f = \w_b \equiv w$, due to the fact that the forward/backward diffusion of a polymer chain essentially comes from Brownian fluctuations. This means that, it is the binding of chaperones that rectifies the diffusion of polymer chain to a directional motion on average. In fact, this result remains valid even if the polymer chain undergoes a biased random walk when crossing the membrane. Note that both the binding and the unbinding of chaperones are extremely fast compared to the forward/backward diffusion of a polymer chain. In all calculations below, we always keep the binding/unbinding rate of a chaperone at least 2 orders of magnitude higher than the forward/backward diffusion rate.
\begin{figure}[h]
	\centering
	\includegraphics[width=0.8\textwidth]{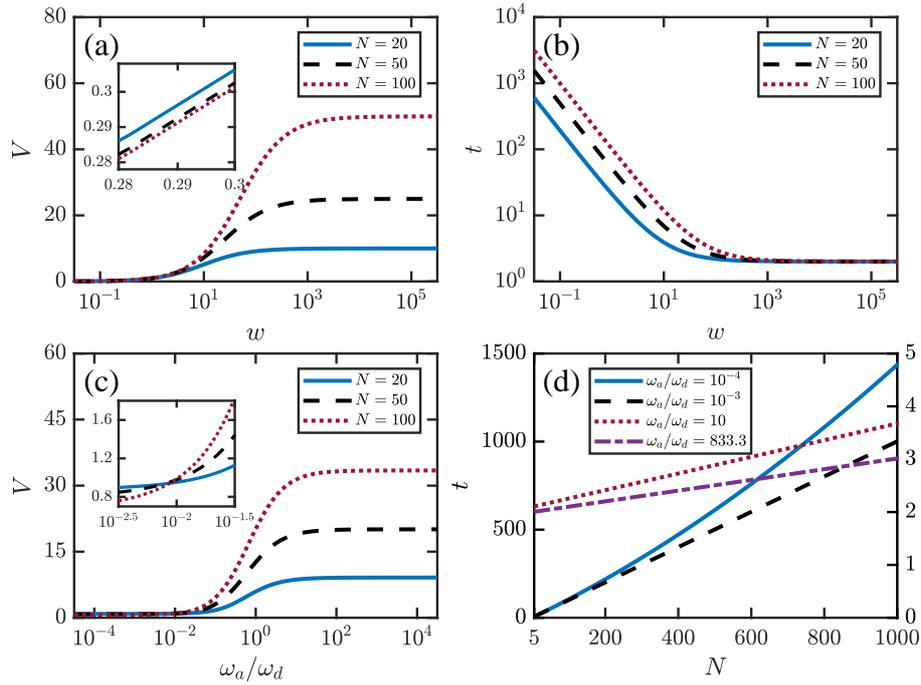}
	\caption{Dependence of  (a) mean velocity $V$ and  (b) mean dwell time $t$ on forward/backward diffusion rate $w$, where the fast binding rate $\w_a = 1000\times10^{8} \text{ s}^{-1}$ and the unbinding rate $\w_d = 1.2\times10^{8} \text{ s}^{-1}$, keeping their ratio $\w_d /\w_a = 0.0012$ \cite{Hepp2016Kinetics}.  (c) Mean velocity $V$ as a function of binding/unbinding ratio $\w_a/\w_d$ with the inset displaying the detail in enlarged scale. (d) Mean dwell time $t$ of translocation as a function of the length of polymer chain with different ratio of the binding/unbinding rate. The solid line and the dashed line are plotted corresponding to the left axis, while the dotted line and dash dotted line are to the right axis. In (c) and (d), the diffusion rate is kept $w = 10^2\text{ s}^{-1}$ and the unbinding rate $\w_d$ is kept of order $10^8$. Thus, the binding/unbinding process is more than 2 orders of magnitude faster than the diffusion process. Other parameters used in calculations are $\w_s =\w_e = 1 \text{ s}^{-1}$.}
	\label{Velocity}
\end{figure}

Numerical results on mean translocation velocity $V$ are displayed in \Fref{Velocity}. When trailing binding sites ({\it i.e.}, binding sites of a polymer chain lying in the target region) are occupied with chaperones in large probability, the polymer chain with a larger forward/backward diffusion rate will pass through the pore more rapidly, since its backward motion is hindered by chaperones. So translocation velocity increases with the diffusion rate $w$. This rise, however, will not last indefinitely, since for sufficiently large diffusion rate $w$, the translocation velocity will be limited by the binding/unbinding process of chaperones as well as the finite starting rate $\w_s$ and ending rate $\w_e$. See \Fref{Velocity}(a). With Eqs.\,(\ref{V}, \ref{b0}), it can be mathematically shown that
 \[\lim_{w\rightarrow +\infty} V=   \frac{N}{\widetilde{W}_1 - q^{N-1}K} \,, \quad \textrm{with}\
  \widetilde{W}_1 = \frac{1}{\w_s} +K \left(\frac{\w_a}{\w_d+\w_e} +\frac{2-q}{1-q} \right)\,.\]
 It means that there is an upper limit for mean velocity $V$ as diffusion rate $w$ tends to infinity. See \Fref{Velocity}(a) again. Besides, we notice that a longer polymer chain employs greater mean velocity provided high diffusion rate, since the restriction of finite starting rate $\w_s$ and ending rate $\w_e$ produces less impact on the overall mean velocity of a long polymer chain. On the contrary,
 if the diffusion rate $w$ is pretty low, it is the relatively rapid initiation and termination that guides the translocation. Then,  shorter polymer chains gain higher mean velocity $V$ since that is when initiation along with termination carries more weights. See the inset of \Fref{Velocity}(a).

From another perspective, we consider the mean dwell time $t$ for some translocation, which can be calculated as
  \begin{equation}\label{timet}
  t = \frac{N}{V} = W_1+ W_2\S^{2-N} + W_3(N-2) \, .
  \end{equation}
See Eqs.\,(\ref{V}, \ref{b0}) for the second equality. Here, the numerator $N$ is the length of a polymer chain and the denominator $V$ is the mean velocity of translocation. \Eref{timet} shows that more time is required for the translocation of a longer polymer chain. According to the definition of constant $S$, if the diffusion rate $w$ is low, $S$ increases almost linearly with $w$, while $W_1, W_2, W_3$  are all insensitive to the low diffusion rate $w$. Therefore, \Eref{timet} indicates that the relation between dwell time $t$ and diffusion rate $w$ follows power law approximately, when the polymer chain diffuses at a low diffusion rate $w$. See \Fref{Velocity}(b).

The dependence of mean velocity $V$ on the ratio of binding/unbinding rate of a chaperone is presented in \Fref{Velocity}(c). With high ratio $\w_a/\w_d$, the binding site lying in the target region will be more likely to be occupied with a chaperone, which will, then, help rectify the forward/backward diffusion. That is, high binding/unbinding ratio speeds up the translocation of a polymer chain. The dependence of mean velocity on ratio $\w_a/\w_d$ displays similar performance to that on diffusion rate $w$, as is shown in \Fref{Velocity}(a, c). If the ratio $\w_a/\w_d$ is large enough, mean translocation velocity $V$ increases with the length $N$ of a polymer chain  and vice versa. See \Fref{Velocity}(c) and the inset. This is because the longer a polymer chain is, the less influence is likely to be exerted through the relatively low starting rate $\w_s$ and ending rate $\w_e$. Contrariwise, low ratio $\w_a/\w_d$ leads to the slow translocation of the main body part of a polymer chain, and the relatively large starting rate $\w_s$ as well as ending rate $\w_e$ plays a more active role in promoting the overall mean translocation velocity $V$ of a shorter polymer chain. It means that the mean velocity will decrease with the growth of the polymer length $N$.

Note that, experiments in \cite{Hepp2016Kinetics} give a binding/unbinding ratio, which is approximate to $833.3$. \Fref{Velocity}(c) shows that the velocity $V$ is almost independent of ratio $\w_a/\w_d$ when $\w_a/\w_d\ge833.3$,  implying that the translocation process is limited by the forward/backward diffusion rate $w$ as well as starting rate $\w_s$ and ending rate $\w_e$. Moreover, the plots in \Fref{Velocity}(a, b) show that, for diffusion rate $w$ larger than  $1000 \text{ s}^{-1}$, mean velocity $V$ is also insensitive to the change of rate $w$. Therefore, the translocation of polymer chain is limited merely by the starting and the ending process when $\w_a/\w_d\ge833.3$ and diffusion rate $w\ge 1000 \text{ s}^{-1}$. Obviously, velocity $V$ increases with both rate $\w_s$  and rate $\w_e$, see \Fref{Schematic}(d) or Eqs.\,(\ref{V}, \ref{b0}). Note that, in \Fref{Velocity}(a, b), the ratio $\w_a/\w_d$ is always kept $833.3$.

The dependence of dwell time $t$ on length $N$ of a polymer chain, with different values of ratio $\w_a/\w_d$, is plotted in \Fref{Velocity}(d). One can easily find that $t$ increases almost linearly with $N$, especially for large binding/unbinding ratio $\w_a/\w_d$.  It can been seen in \Eref{timet} as well. Meanwhile, \Fref{Velocity}(d) also shows that dwell time $t$ decreases with the ratio $\w_a/\w_d$, since a polymer chain deserves high mean velocity $V$ provided with great binding/unbinding ratio $\w_a/\w_d$. See \Fref{Velocity}(c) and the definition of $t$ given in \Eref{timet}.

\begin{figure}[h]
	\centering
	\includegraphics[width=0.8\textwidth]{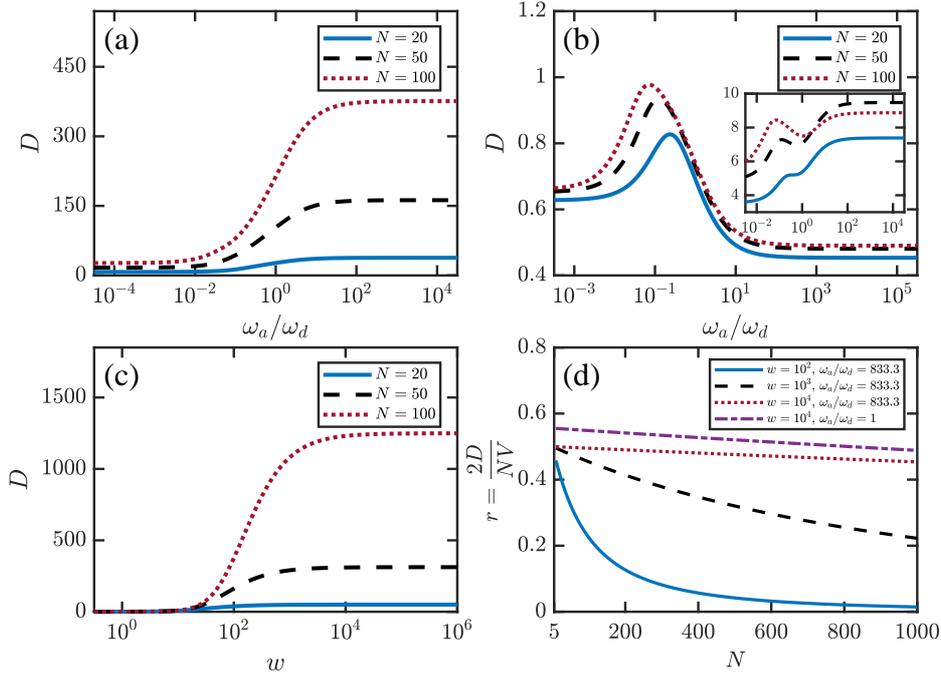}
	\caption{Effective diffusion constant $D$ as a function of the ratio $\w_a/\w_d$ is displayed in (a) and (b). The diffusion rate used in (a) is $w=10^2 \text{ s}^{-1}$, while calculations in (b) are carried out with lower diffusion rates. $w = 1 \text{ s}^{-1}$ in (b) and $w = 10 \text{ s}^{-1}$ in inset. (c) Dependence of $D$ on the diffusion rate $w$, with ratio $\w_a/\w_d = 833.3$ as is given in \cite{Hepp2016Kinetics}. (d) The randomness, $r\equiv 2D/NV$, as a function of the length $N$ of a polymer chain. Several cases with different values of diffusion rate $w$ and ratio $\w_a/\w_d$ are considered. Detailed instructions on parameters are listed in the legend. Similar to \Fref{Velocity}, $\w_d$ is held of order $10^{8} \text{ s}^{-1}$ to make the binding/unbinding process of a chaperone much faster than the forward/backward diffusion process of a polymer chain. Other parameters used in calculations are $\w_s = 1 \text{ s}^{-1}$ and $\w_e =1 \text{ s}^{-1}$. }
	\label{Diffusion}
\end{figure}

The discussion above on mean velocity $V$ provides a chief description of the dynamics of polymers' translocation across membrane, but it is not sufficient for many biologically relevant cases, especially when the polymer chain is not long. In these situations, effective diffusion constant $D$, or say dispersion, plays an important role. The plots in  \Fref{Diffusion}(a) show that, for high diffusion rate $w$, diffusion constant $D$ increases with both the ratio $\w_a/\w_d$ and the polymer length $N$. At the same time, diffusion constant $D$ tends to its lower limit or upper limit monotonically as the ratio $\w_a/\w_d$ tends to 0 or infinity, respectively. However, something interesting happens as diffusion rate $w$ goes down approaching the starting rate $\w_s$ and ending rate $\w_e$. The plots in \Fref{Diffusion}(b) show that, for such cases, $D$ will no longer change monotonically with rate $\w_a/\w_d$. Neither will it increase with the polymer length $N$. With a particular value of ratio $\w_a/\w_d= 833.3$, as is given in \cite{Hepp2016Kinetics}, the dependence of $D$ on diffusion rate $w$ is plotted in \Fref{Diffusion}(c). As expected, $D$ increases with $w$ and tends to a limit rising with the polymer length $N$.

Compare \Fref{Velocity}(a, c) with \Fref{Diffusion}(a, c), and one can find that mean velocity $V$ and effective diffusion constant $D$ show almost the same behavior. Both of them increase significantly when the parameter, either forward/backward diffusion rate $w$ or binding/unbinding ratio $\w_a/\w_d$, varies within some range. With the parameter varying beyond this range, however, mean velocity $V$ and effective diffusion constant $D$ both stay almost constant. The reason is that the dynamics of translocation process is regulated jointly by several factors. Besides diffusion rate $w$ along with the ratio $\w_a/\w_d$, these factors include starting rate $\w_s$ and ending rate $\w_e$. See \Fref{Schematic}. When the corresponding parameter, rate $w$ or $\w_a/\w_d$, varies within an appropriate range, the translocation process of a polymer is dominated by its diffusion process and the binding/unbinding process of a chaperone, and the mean velocity or diffusion constant changes significantly with the parameter, rate $w$ or ratio $\w_a/\w_d$. Otherwise, the dynamics of translocation will be limited by rate $\w_s$ and $\w_e$, and therefore is almost independent of the rate $w$ and ratio $\w_a/\w_d$. In other words, the translocation of a polymer chain across membrane consists of several kinetics, namely initiation, termination, binding/unbinding of chaperones and forward/backward diffusion of a polymer chain. For a given set of parameters, the overall translocation process may be limited by only one or some of them, with others fast enough to be neglected. But the limit process may be switched from one to another, with the change of one or some parameter values

It is also attractive to compare the degree of fluctuations of this stochastic process. An important dimensionless function to evaluate this quantity is randomness, which is defined as \cite{Kolomeisky2005Dynamic,Svoboda1994Fluctuation}
\[r = \frac{2D/N^2}{V/N} = \frac{2D}{NV} \,.\]
The plots in \Fref{Diffusion}(d) show that, the randomness $r$ always decreases with the polymer length $N$. With the ratio $\w_a/\w_d= 833.3$, which is consistent with \cite{Hepp2016Kinetics}, $r$ increases with the forward/backward diffusion rate $w$. Our results also suggest a decrease of  $r$ with the number of binding sites, {\it i.e.}, $N$. A rapid decay in randomness $r$ can be observed on the condition of a low diffusion rate $w$, whereas the length of a polymer reduces $r$ slightly, but almost linearly, as long as the diffusion rate $w$ is high.

\section{Conclusions and Remarks}\label{sec5}
The \lq\lq Brownian Ratchet"  builds a general framework describing how diffusive motion is  rectified by chemical energy.  Particularly, the chaperone-assisted translocation of polymer chains across membrane constitutes a fine example of how directed motion can emerge from random diffusion, where the binding and unbinding of chaperones plays a prominent role.

In this paper, by mapping the process of translocation across membrane into the discrete motions of a membrane on the polymer chain, a theoretical model is presented, where the probability of finding membrane at some site of polymer chain is governed by usual master equations. With this model, the mean velocity and effective diffusion constant of translocation process can be derived explicitly, allowing us to discuss the dynamics of translocation across membrane much more efficiently.

Based on the exact expressions, detailed discussions on particular cases are presented, where the polymer chain is assumed to diffuse freely around the nanopore embedded in membrane if there is no chaperone molecule. Our results show that both the increase of forward/backward diffusion rate $w$ of polymer chain and the rise in binding/unbinding rate ratio $\w_a/\w_d$ of chaperones raise the mean translocation velocity monotonically to a polymer length dependent limit. With large diffusion rate $w$ or great ratio $\w_a/\w_d$, the mean velocity increases with the length of polymer chain, while opposite performance can be observed when $w$ or $\w_a/\w_d$ stays at a low level. The effective diffusion constant also increases with the diffusion rate $w$, but it will exhibit complicated properties when regarded as a function of ratio $\w_a/\w_d$. Meanwhile, the randomness decreases monotonically with the length of a polymer chain.

For most of biological processes in living cells, mean velocity together with effective diffusion constant is usually sufficient to describe the basic dynamic properties. Nevertheless, higher orders  moments, or say higher cumulants, can be obtained as well, simply via the same methods in this paper. Meanwhile, all of the transition rates are assumed to be independent of the site of a polymer chain in our discussion. But in a promoted model with site-dependent transition rates, explicit expressions for mean velocity and effective diffusion constant can still be obtained sketching the key idea of this study.


 \begin{appendix}
\section{Derivation of \Eref{D}}\label{app}
Since the final expression for diffusion constant $D$ does not depend on the undetermined constant $T_0$, we let $T_0 = 0$, which indicates both $X_k$s and $\widetilde{X}_k$s equal to zero. At the same time, $Y_k$ follows \Eref{Yk0} and $\Y_k$ follows \Eref{Yk1}.

 In the stationary-state limit,  \Eref{D1}, \Eref{D2} and \Eref{D3} turn to
 \begin{equation}\label{AD1}
 \frac{1}{2}D_1 =\w_f Y_1 +\w_{f_0}\sum_{k=2}^{N-1}Y_k+\w_{f_1}\sum_{k=2}^{N-1}\Y_k -\w_b \sum_{k=2}^{N} Y_k \,,
 \end{equation}
 \begin{equation}\label{AD2}
 \frac{1}{2}D_2 =\frac{1}{2}\left[\w_s b_0 +\w_f b_1 +\w_{f_0} \sum_{k=2}^{N-1} b_k + \w_{f_1} \sum_{k=2}^{N-1} \b_k  +\w_b\sum_{k=2}^{N} b_k
 \right]\,,
 \end{equation}
 and
  \begin{equation}\label{AD3}
- \frac{1}{2}D_3 = - A\left[Y_1+\sum_{k=2}^N(Y_k +\Y_k)\right]\, .
 \end{equation}
 Take \Eref{Tk1} into \Eref{AD1},
   \begin{equation}\label{A2}
	   \begin{split}
			\frac{1}{2}D_1 =&(\w_f Y_1-\w_b Y_2) + \sum_{k=2}^{N-1}(SY_k-RY_{k+1})-\frac{\w_{f_1}}{\w_d +\w_{f_1}} \sum_{k=2}^{N-1}\a_k -(1-q)\w_b\frac{\w_{f_1}}{\w_d +\w_{f_1}}\sum_{k=2}^{N-1}b_{k+1}\, ,\\
			=&  -\frac{\w_{f_1}}{\w_d +\w_{f_1}} \sum_{k=2}^{N-1}\a_k -(1-q)\w_b\frac{\w_{f_1}}{\w_d +\w_{f_1}}\sum_{k=2}^{N-1}b_{k+1}+\sum_{k=1}^{N-1} r_k\, .
	   \end{split}
  \end{equation}
With \Eref{rk}, we have
\begin{equation}\label{A3}
		\sum_{k=1}^{N-1}r_k
		= -\frac{\w_{d}}{\w_d +\w_{f_1}}\sum_{k=2}^{N-1}\a_k -R\sum_{k=2}^{N-1}b_{k+1}-\w_b b_2+\w_sb_0 \sum_{k=1}^{N-1}k-\sum_{k=1}^{N-1}\sum_{i=1}^{k}a_i -\sum_{k=2}^{N-2}\sum_{i=2}^{k}\a_i \, .
\end{equation}
Note that the summation of the first term in \Eref{A2} and that in  \Eref{A3} is
\begin{equation*}
-\frac{\w_{f_1}}{\w_d +\w_{f_1}} \sum_{k=2}^{N-1}\a_k-\frac{\w_d}{\w_d +\w_{f_1}}\sum_{k=2}^{N-1}\a_k =-\sum_{i=2}^{N-1}\a_i \, ,
\end{equation*}
and the summation of terms related to $b_k$ in these two equations is shown as
\begin{equation*}
 -(1-q)\w_b\frac{\w_{f_1}}{\w_d +\w_{f_1}}\sum_{k=2}^{N-1}b_{k+1}-R\sum_{k=2}^{N-1}b_{k+1}-\w_b b_2+\w_sb_0 \sum_{k=1}^{N-1}k
=-\w_b\sum_{k=2}^N b_k +\frac{1}{2}N(N-1)\w_s b_0 \,.
\end{equation*}
Then
\begin{equation}\label{A6}
\frac{1}{2}D_1 = -\sum_{k=1}^{N-1}s_k -\sum_{k=2}^{N-1}\widetilde{s}_k -\w_b\sum_{k=2}^Nb_k +\frac{1}{2}N(N-1)\w_s b_0 \, ,
\end{equation}
where
\[s_k =\sum_{i=1}^k a_i\,, \quad \widetilde{s}_k =\sum_{i=2}^k\a_i \,.\]

For \Eref{AD2},
 \begin{equation}\label{A7}
 \begin{split}
 \frac{1}{2}D_2 =&\frac{1}{2}\left[
 \w_s b_0 +\w_f b_1 +\w_{f_0} \sum_{k=2}^{N-1} b_k + \w_{f_1} \sum_{k=2}^{N-1} \b_k  -\w_b\sum_{k=2}^{N} b_k+2\w_b\sum_{k=2}^{N} b_k
 \right]\\
 =&\w_b\sum_{k=2}^{N} b_k +\frac{1}{2}V\,.
 \end{split}
\end{equation}
With \Eref{V}, the summation of \Eref{A6}, \Eref{A7} and \Eref{AD3} can be written as
\begin{equation*}
\begin{split}
D =& \frac{1}{2}(D_1+D_2-D_3)\\
=& -\sum_{k=1}^{N-1}s_k -\sum_{k=2}^{N-1}\widetilde{s}_k  +N\frac{V}{2}-V \left[Y_1+\sum_{k=2}^N(Y_k +\Y_k)\right] \, ,
\end{split}
\end{equation*}
which is exactly \Eref{D}.
\end{appendix}


\end{document}